\documentclass[twocolumn,amssymb,amsmath,nobibnotes,nofootinbib,showpacs,aps,prb]{revtex4-1}
\usepackage[dvips]{graphicx,color}
\usepackage{caption}
\begin{document}
\title{High-temperature ferroelectric order and magnetic field-cooled effect driven magnetoelectric coupling in R$_2$BaCuO$_5$ (R= Er, Dy, Sm)}
\author{A. Indra$^{1,2}$}
\author{S. Mukherjee$^{1}$}
\author{S. Majumdar$^1$}
\author{O. Gutowski$^3$}
\author{M. v. Zimmermann$^3$}
\author{S. Giri$^1$}
\affiliation{$^1$School of Physical Sciences, Indian Association for the Cultivation of Science, Jadavpur, Kolkata 700032, India}
\affiliation{$^2$Department of Physics, Srikrishna College, Bagula, Nadia, W. B., 741502, India}
\affiliation{$^3$Deutsches Elektronen-Synchrotron DESY, Notkestr. 85, 22603 Hamburg, Germany}

\begin{abstract}
The high-temperature ferroelectric order and a remarkable magnetoelectric effect driven by the magnetic field cooling are reported in R$_2$BaCuO$_5$ (R = Er, Dy, Sm) series. The ferroelectric (FE) orders are observed at much higher temperatures than their magnetic orders for all three members. The value of FE Curie temperature ($T_{FE}$) is considerably high as $\sim$ 235 K with the polarization value ($P$) of $\sim$ 1410 $\mu$C/m$^2$ for a 4 kV/cm poling field in case of Er$_2$BaCuO$_5$, whereas the values of $T_{FE}$ and $P$ are also promising as $\sim$ 232 K and $\sim$ 992 $\mu$C/m$^2$ for Dy$_2$BaCuO$_5$, and $\sim$ 184 K and $\sim$ 980 $\mu$C/m$^2$ for Sm$_2$BaCuO$_5$. The synchrotron diffraction studies of Dy$_2$BaCuO$_5$ confirm a structural transition at $T_{FE}$ to a polar $Pna2_1$ structure, which correlates the FE order. An unusual magnetoelectric coupling is observed below the R order for Er and Dy compounds and below the Cu order for Sm compound, when the pyroelectric current is recorded only with the magnetic field both in heating and cooling cycles i.e. typical magnetic field cooled effect. The magnetic field cooled effect driven emergence of polarization is ferroelectric in nature, as it reverses due to the  opposite poling field. The unexplored R$_2$BaCuO$_5$ series attracts the community for large $T_{FE}$, high $P$ value, and strange magnetoelectric consequences.  
\end{abstract}
\pacs{75.85.+t, 75.80.+q, 77.80.-e}
\maketitle
\section{Introduction}
Multiferroics, where ferroelectric order coexist with the long range magnetic order, attract special attention for the magnetoelectric (ME) cross coupling.\cite{spaldin,fie,cheo} In addition to the fundamental interest on the origin of coexisting ME orders in a chemically single phase compounds, multiferroics are extremely promising for the applications.\cite{scott,spaldin1,ramesh1,spaldin2} For the practical applications, occurrence of ME orders associated with the strong ME coupling is highly desirable, which is still missing except for very few promising inorganic materials like BiFeO$_3$,\cite{wang,cher} CuO,\cite{kimura_NM} and Sr$_3$Co$_2$Fe$_{24}$O$_{41}$.\cite{soda} Herein, occurrence of the ferroelectric (FE) order is revealed at significantly high temperature in an unexplored R$_2$BaCuO$_5$ (R= Er, Dy, Sm) series. For example, the FE order is observed around $\sim$ 235 K associated with a large value of electric polarization ($P$) of $\sim$ 1410 $\mu$C/m$^2$ for a 4 kV/cm poling field (E) in case of Er$_2$BaCuO$_5$. Furthermore, an intriguing magnetic field cooled effect driven strong ME coupling is observed below the magnetic ordering temperature in the R$_2$BaCuO$_5$ series.

\begin{figure}[t]
\centering
\includegraphics[width = \columnwidth]{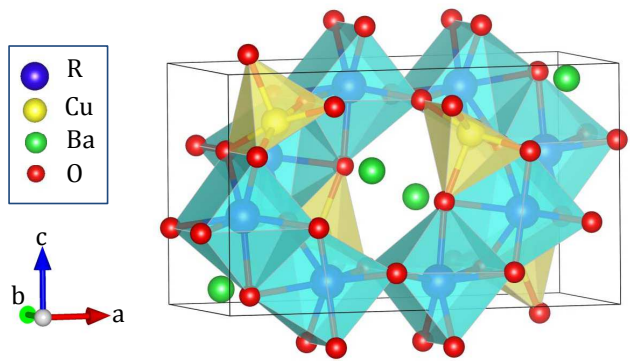}
\caption {Direct linkage of RO$_7$ polyhedra and linkage through the CuO$_5$ polyhedra via O atoms. Atomic positions are described in the figure.} 
\label{str}
\end{figure}

 The R$_2$BaCuO$_5$ series of compounds crystallize in the $Pnma$ (Z = 4) space group, where the copper ions are occupied in the distorted square pyramids (CuO$_5$) and are connected by the RO$_7$ polyhedra.\cite{michel,what,hunt,sal} Direct linkages between RO$_7$ and linkage through CuO$_5$ are depicted in Fig. \ref{str}. It has been observed that the lattice parameters and volume of the unit cell decrease linearly as a function of the lanthanide ionic radius starting from Sm$^{3+}$ to Lu$^{3+}$.\cite{sal} The short-range force constant model was used for interpreting the results of Raman and infrared wavenumbers in the orthorhombic phase of R$_2$BaCuO$_5$ (R = Y, Ho, Gd), which was consistent with the orthorhombic phase of $Pnma$ space group.\cite{gupta} The heat capacity\cite{mosh} and magnetization results\cite{sal1,lev} confirmed two magnetic transitions corresponding to antiferromagnetic (AFM) Cu$^{2+}$ and R$^{3+}$ orders, respectively. The EPR spectra were recorded for few members of R$_2$BaCuO$_5$ series, where strong exchange coupling was suggested between Cu$^{2+}$ and R$^{3+}$ ions.\cite{mez} In fact, significant Cu$^{2+}$ and R$^{3+}$ exchange interaction and large magnetic anisotropy were proposed from the magnetic and spectroscopic investigations for R = Dy and Ho.\cite{baran} 
 High resolution optical absorption spectra of an intrinsic R$^{3+}$ ion and Er$^{3+}$ probe further confirmed the magnetic ordering temperatures and proposed that the magnetic structures of the copper sub-system are same for all the members of R$_2$BaCuO$_5$ (R = Sm, Eu, Tm, Yb, Lu).\cite{pau} Furthermore, the optical spectra in the "green phase" of Dy$_2$BaCuO$_5$ proposed the first order magnetic phase transition involving Cu and Dy moments.\cite{popo} Neutron diffraction studies have been performed for Nd and Gd compounds at the R-site. The magnetic structure of Nd$_2$BaCuO$_5$ was suggested on the basis of a propagation vector k = [0, 0, 1/2] below the N\'{e}el temperature of 7.8 K, where the Cu$^{2+}$ and Nd$^{3+}$ magnetic moments were aligned along the crystallographic $c$-axis and in the $ab$-plane, respectively.\cite{puche} Recent neutron results on isotopically substituted $^{160}$Gd$_2$BaCuO$_5$ proposed an incommensurate magnetic structure with a propagation vector k = [0 0 1/2 - $\delta$] below 12 K, followed by a commensurate structure with a k = [0 0 1/2] propagation vector at 5 K, which was accompanied with a Gd spin re-orientation and a magnetostructural transition.\cite{ovsy} The considerable magnetic refrigeration with a maximum magnetic entropy change of -$\Delta$S$_M \approx$ 10.4 J/kg K at N\'{e}el temperature and refrigerant capacity of $\approx$ 263 J/kg were suggested for Ho$_2$BaCuO$_5$.\cite{zhang} 

Recently, the spin-chain compounds R$_2$BaNiO$_5$\cite{sharma,basu,basut,singh,pal} and R$_2$BaCoO$_5$\cite{upad} attract special attentions for the multiferroic order, proposing different origins behind the occurrence of  ferroelectric order. In this study we also observe the ferroelectric order in the unexplored R$_2$BaCuO$_5$ (R= Er, Dy, Sm) series. Intriguingly, ferroelectricity in all the members occurs at much higher temperatures than the magnetic ordering temperatures. The values of ferroelectric (FE) ordering temperatures ($T_{FE}$) are close to room temperature as $\sim$ 235 K and $\sim$ 232 K for Er$_2$BaCuO$_5$ (EBCO) and Dy$_2$BaCuO$_5$  (DBCO), respectively, whereas the value is $\sim$ 184 K for Sm$_2$BaCuO$_5$ (SBCO). The values of $P$ are high as $\sim$ 1410, $\sim$ 992, and $\sim$ 980 $\mu$C/m$^2$ for E = 4 kV/cm in case of EBCO, DBCO, and SBCO, respectively. An interesting ME coupling driven by the magnetic field cooled (FC) effect is observed below the magnetic ordering temperatures for all the members of the series of compounds. The magnetic FC effect driven increase of $P$ is similar for EBCO and DBCO, which is different for SBCO. These FC effects are strongly correlated to the MD effect as well as magnetization curves of R$_2$BaCuO$_5$. The synchrotron diffraction studies of DBCO over a wide temperature range of 10-300 K confirm a structural transition to a polar structure of $Pna2_1$ space group from the centrosymmetric $Pnma$ structure at 232 K, around which the ferroelectricity appears. We suggest that the appearance of ferroelectricity correlates the structural transition to a polar structure.

\begin{figure}[t]
\centering
\includegraphics[width = \columnwidth]{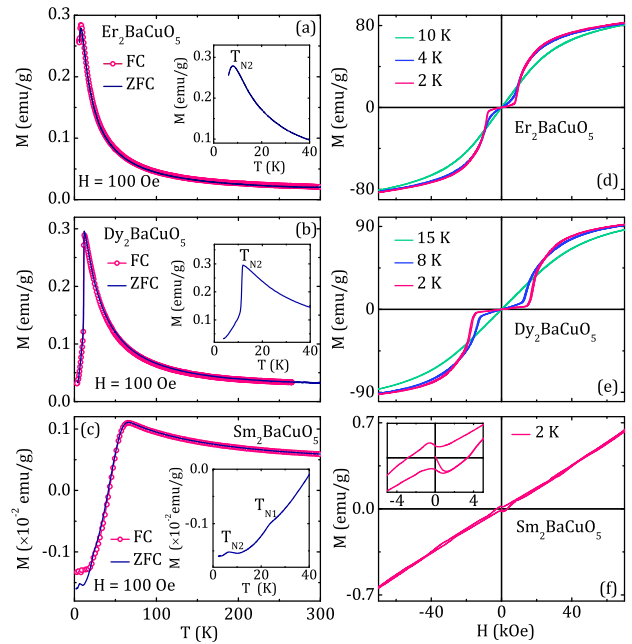}
\caption {Temperature dependences of FC-ZFC magnetization curves for (a) EBCO, (b) DBCO, and (c) SBCO recorded with  100 Oe. The insets highlight the low-$T$ region of each $M(T)$ curve. Magnetization curves for (d) EBCO, (e) DBCO, and (f) SBCO at selected temperatures. Inset of (f) highlights the low filed region of the $M(H)$ curve.}  
\label{mag}
\end{figure}

\section{Experimental details}

Polycrystalline R$_2$BaCuO$_5$ (R= Er, Dy, Sm) compounds are prepared using the solid-state reaction.\cite{sal} The single phase chemical composition is confirmed by the x-ray diffraction studies at room temperature recorded in a PANalytical x-ray diffractometer (Model: X’ Pert PRO) using the Cu K$\alpha$ radiation. The single-phase chemical composition is further checked by the synchrotron x-ray diffraction studies recorded with a wavelength of 0.1259 \AA~ (98 keV) at the P07 beamline of PETRA III, Hamburg, Germany in the temperature range of 10-300 K. The synchrotron powder diffraction data are analyzed using the Rietveld refinement with the commercially available MAUD and FullProf softwares. Powder sample, pressed into a pellet, is used for the dielectric measurements using a E4980A LCR meter (Agilent Technologies, USA) equipped with a PPMS-II system of Quantum Design. The pyroelectric current ($I_p$) is recorded in an electrometer (Keithley, model 6517B) at a constant temperature sweep rate. The $I_p$ is integrated over time for obtaining the spontaneous electric polarization. The poling electric field is applied during cooling processes and the measurements are carried out in the warming mode with a zero electric field. Before measurement of $I_p$, the electrical connections are short circuited and waited for a long time. In all the measurements the electrical contacts are fabricated using an air drying silver paint. The temperature dependence heat capacity ($C_p$) is measured with a PPMS-I system of Quantum Design. Magnetization is measured in a commercial magnetometer of Quantum Design (MPMS, evercool) both in zero-field-cooled (ZFC) and FC protocols. In case of ZFC and FC conditions sample is cooled in zero magnetic field ($H$) and non-zero $H$ and measurements are carried out in the warming mode with a non-zero $H$.  

\begin{figure}[t]
\centering
\includegraphics[width = \columnwidth]{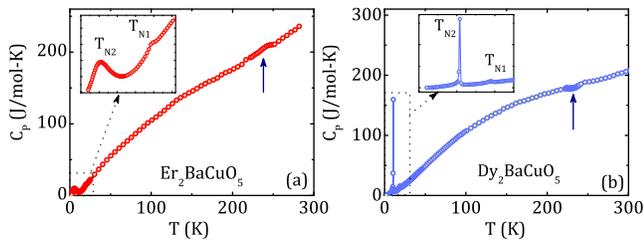}
\caption {The $T$ variations of heat capacity ($C_p$) for (a) EBCO and (b) DBCO. The inset of all figures show the highlighted region of $C_p$.The arrows in the high temperature region depict the FE ordering temperature.} 
\label{Cp}
\end{figure}

\section{Experimental results and discussions}

Thermal variations of the ZFC-FC magnetization curves recorded at 100 Oe are displayed in Figs. \ref{mag}(a), \ref{mag}(b), and \ref{mag}(c) for EBCO, DBCO, and SBCO, respectively. In accordance with the reported results, a maximum representing the antiferromagnetic (AFM) ordering of the R ion ($T_{N2}$) is observed at $\sim$ 5 K and 11 K for EBCO and DBCO, respectively.\cite{lev,mez} The insets of Figs. \ref{mag}(a) and \ref{mag}(b) highlight the low temperature transitions. A sharp fall of magnetization below $\sim$ 11 K rather proposes the first order magnetic phase transition, as originally suggested from the optical studies in DBCO.\cite{popo} Any definite signature of similar first order magnetic phase transition is missing for EBCO in the current results because of the limitation of the measurement facility below 2 K. Nature of the ZFC-FC magnetization curves for SBCO is quite different from the other two compounds. As shown in Fig. \ref{mag}(c), the $M(T)$ decreases rapidly below $\sim$ 100 K and becomes negative below 50 K. Similar $M(T)$ curves have been noted for the different systems proposing different origins.\cite{yusuf} The interplay between different temperature dependences of the rare earth and $3d$ moments were correlated to the negative magnetization in the $3d$-$4f$ oxide systems. Along with the negative magnetization, a weak magnetization is noted in SBCO than other two compounds. This might be attributed to the close ordered moments of Sm$^{3+}$ and Cu$^{2+}$ ions with an antiparallel exchange coupling. As depicted in the inset of Fig. \ref{mag}(c), two anomalies are observed around $\sim$ 23 K ($T_{N1}$) and 6 K ($T_{N2}$), which represent the AFM ordering temperatures of Cu$^{2+}$ and Sm$^{3+}$, respectively.\cite{lev,mez} 

The magnetic hysteresis loops are recorded below the transition temperatures as depicted in Figs. \ref{mag}(d), \ref{mag}(e), and \ref{mag}(f) for EBCO, DBCO, and SBCO, respectively. The EBCO and DBCO show field induced transition from AFM to a ferromagnetic (FM) state below $T_{N2}$. The field induced transition shifts toward lower magnetic field with increasing temperature and vanishes above $T_{N2}$. The nature of $M(H)$ curve is different for SBCO. An almost linear magnetization curve is observed at 2 K, as depicted in Fig. \ref{mag}(f), which is consistent with the proposed AFM order. The nature of the $M(H)$ curve in the low field region is highlighted in the inset of Fig. \ref{mag}(f). A negative magnetization is observed in the low positive magnetic field up to 3 kOe. This negative magnetization below 3 kOe is consistent with the negative $M(T)$ curve in the low temperature recorded at 100 Oe.

\begin{figure}[t]
\centering
\includegraphics[width = \columnwidth]{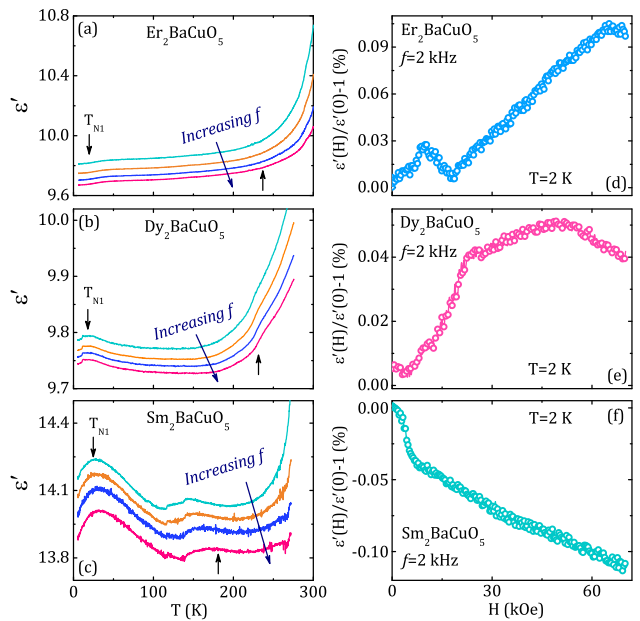}
\caption {(Color online) The $T$ variations of real the component of dielectric permittivities ($\epsilon^{\prime}$) at different frequencies ($f$) for (a) EBCO, (b) DBCO, and (c) SBCO. The arrows in the high temperature region depict the FE ordering temperature. The magnetodielectric response defined as, $\epsilon^{\prime} (H)/\epsilon^{\prime} (0)$ - 1, with $H$ for (d) EBCO, (e) DBCO, and (f) SBCO.} 
\label{die}
\end{figure}

To find out the onset of Cu$^{2+}$ ordering, which is not evident in the $M(T)$ curve of EBCO and DBCO, we incorporate the specific heat capacity ($C_P$) measurements. Figures \ref{Cp}(a) and \ref{Cp}(b) show the $C_P(T)$ of EBCO and DBCO, respectively. The insets of the two figures highlight the low temperature region of $C_P(T)$. A second anomaly ($T_{N1}$) is evident around $\sim$ 19 and $\sim$ 18 K for EBCO and DBCO, respectively, which is associated with the strong signature of $T_{N2}$ at low temperature.\cite{lev, mosh} In addition to the low temperature results, the weak anomalies, as indicated by the arrows in Figs. \ref{Cp}(a) and \ref{Cp}(b) for EBCO and DBCO, respectively, are also observed around $\sim$ 235 K and $\sim$ 232 K. We note that the onset of the polar order is noted close to those temperatures for EBCO and DBCO, which is discussed elsewhere in this article.

\begin{figure*}[t]
\centering
\includegraphics[width = 16cm]{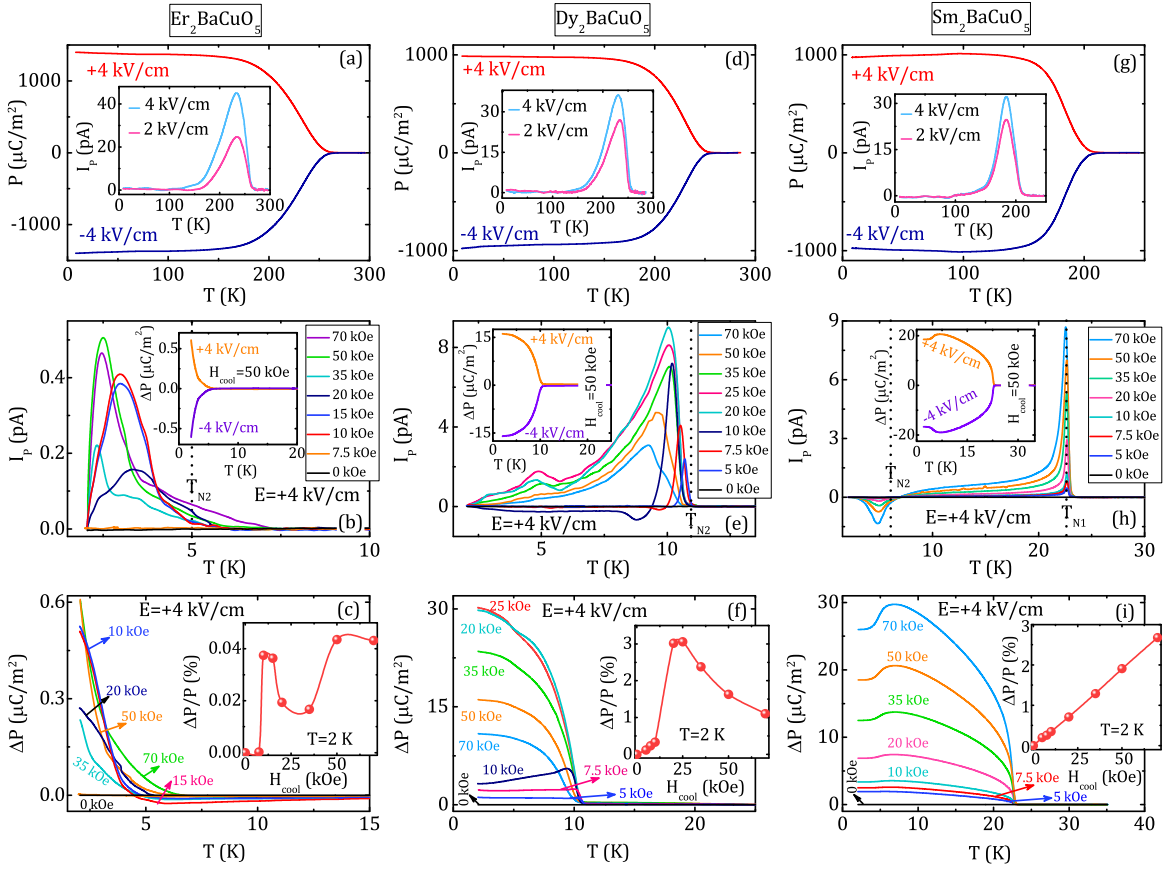}
\caption {The $T$ variations of polarization ($P$) estimated at opposite poling fields ($E = \pm$ 4 kV/cm) of (a) EBCO, (d) DBCO, and (g) SBCO. Insets show $T$ variations of pyroelectric current ($I_p$) at two different $E$. The $T$ variations of $I_p$ curves for different magnetic cooling fields ($H_{cool}$) of (b) EBCO, (e) DBCO, and (h) SBCO. The positions of $T_{N1}$ and $T_{N2}$ are given by the vertical broken lines. Insets shows the $\Delta P(T)$ curves with $H_{cool}$ = 50 kOe and $E = \pm$ 4 kV/cm. The $T$ variations of $\Delta P(T)$ curves for different $H_{cool}$ of (c) EBCO, (f) DBCO, and (i) SBCO. Insets of the corresponding figures show the variations of $\Delta P/P$ with $H_{cool}$.} 
\label{pyro}
\end{figure*}
\begin{figure}[t]
\centering
\includegraphics[width = \columnwidth]{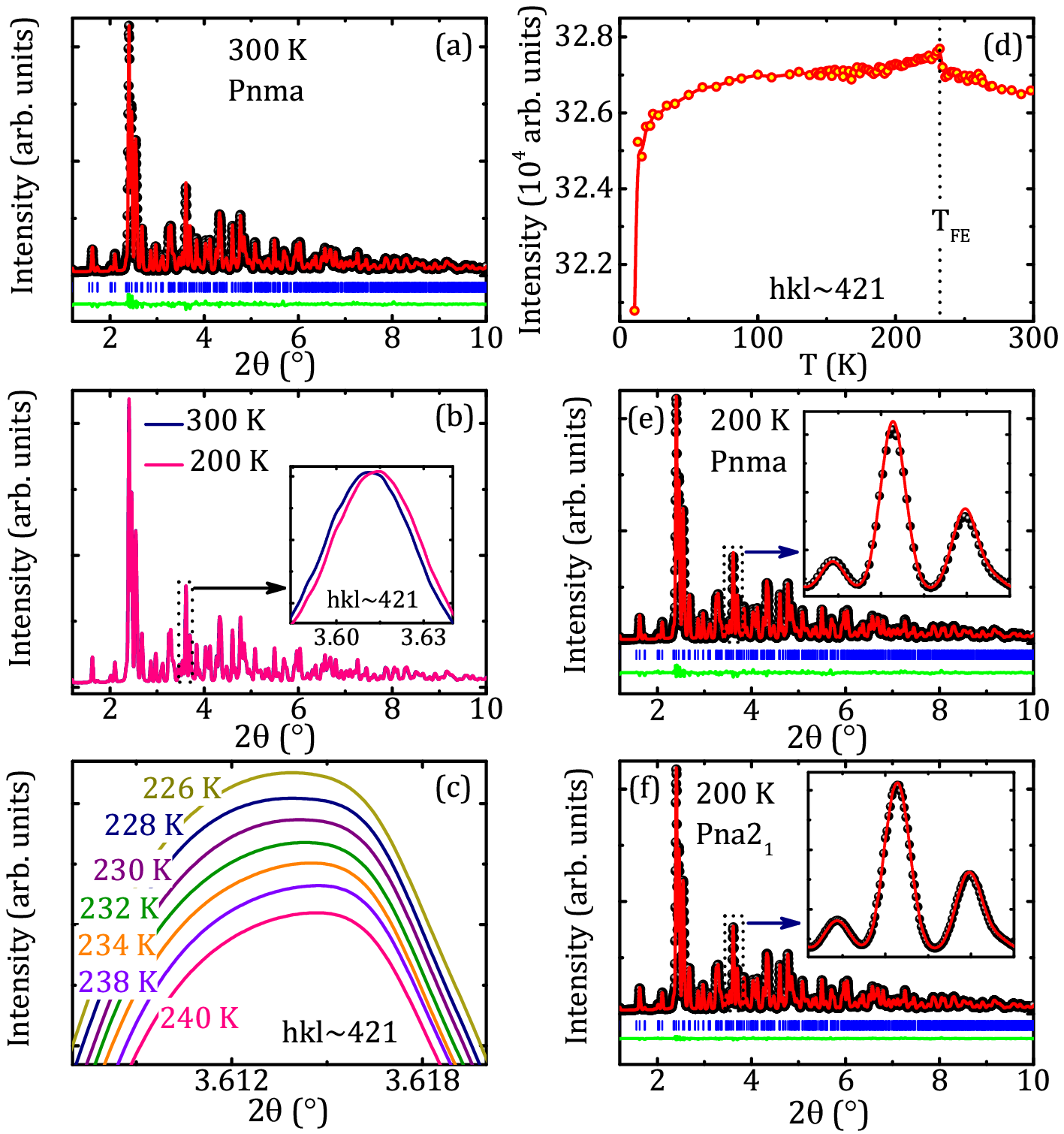}
\caption {(a) The Rietveld refinement of the synchrotron powder diffraction pattern (symbols) at 300 K for DBCO. The solid curve demonstrates the fit. (b) Synchrotron diffraction patterns at 200 and 300 K. Inset highlights the (421) peak. (c) The (421) peak positions with $T$ in a small interval between 226 - 240 K. (d) $T$ variation of the integrated intensity of (421) peak. Rietveld refinements of the diffraction patterns at 200 K using  (e) $Pnma$ and (f) $Pna2_1$ space group. Insets further highlight the quality of the refinements in a small 2$\theta$ region.}
\label{xrd}
\end{figure}

The dielectric permittivity ($\epsilon$) is recorded at different frequencies ($f$) by varying $T$ for all three compounds. Figures \ref{die}(a), \ref{die}(b), and \ref{die}(c) depict thermal variations of the real components ($\epsilon^{\prime}$) of $\epsilon$ for EBCO, DBCO, and SBCO, respectively. The $\epsilon^{\prime}$($T$) plots demonstrate a weak anomaly or a change of slope in the $\epsilon^{\prime}$($T$) curve in the high temperature region, similar to that observed in the $C_P$($T$) curve, which is indicated by the arrows in the figures. Here, the arrows indicate the onset of the spontaneous electric polarization. The details of which are discussed later. A peak or anomaly is also observed in the low-$T$ region, as indicated by the another arrows in the figures, around which the $T_{N1}$ is observed for all the three compounds, indicating the ME coupling. 

The magnetodielectric (MD) effects are recorded at low temperature for all the compounds. The MD effects, defined as $\epsilon^{\prime} (H)/\epsilon^{\prime} (0)$ - 1, are depicted with $H$ in Figs. \ref{die}(d-f) for EBCO, DBCO, and SBCO. The $\epsilon^{\prime} (H)$ and $\epsilon^{\prime} (0)$ represent the $\epsilon^{\prime}$ with $H$ and $H$ = 0, respectively. As depicted in Figs. \ref{die}(d) and \ref{die}(e), the $H$ variations of MD effect for EBCO and DBCO are correlated to the observed magnetization curves at 2 K. The changes in the MD(\%) curve at 2 K are rapid around $\sim$ 7.5 and $\sim$ 20 kOe, around  which the $H$ induced transitions are noticed in the magnetization curves at 2 K for EBCO and DBCO, respectively. With the further increase of $H$ the MD(\%) effect of EBCO shows a peak around $\sim$ 8 kOe, above which it decreases showing a 'dip', and finally it increases systematically above $\sim$ 18 kOe. The MD(\%) effect of DBCO is different from the result of ECBO. It shows a change of slope around $\sim$ 23 kOe, above which it further increases displaying a maximum around $\sim$ 50 kOe. The MD effect of SBCO is different from the rest of two. Initially, it decreases rapidly, around which a negative $M(H)$ with the increase of low $H$ up to $\sim$ 3 kOe is observed in the magnetization curve at 2 K. With further increase in $H$, the MD(\%) decreases monotonically at 2 K.   

In order to confirm the occurrence of spontaneous polar order, the pyroelectric currents ($I_p$) are recorded with $T$ for all the samples in different conditions. A peak in $I_p$($T$) is observed for all the samples, as evident in the insets of Figs. \ref{pyro}(a), \ref{pyro}(d), and \ref{pyro}(g) with poling fields ($E$) of 2 and 4 kV/cm for EBCO, DBCO, and SBCO, respectively. The peaks of $I_p$($T$) curves appear at 235 K, 232 K and 184 K for EBCO, DBCO, and SBCO, respectively. To confirm the peaks in $I_p(T)$, appear due to genuine occurrence of the polar order, the $I_p(T)$ are recorded for different poling temperatures, which are above and below the peak-temperatures for all three compounds. In all the measurements the definite signatures of the peaks are always observed at the same temperature, as evident in Fig. S1 of the Supplemental Material, pointing genuine occurrence of the polar order at the peaks in $I_p(T$) curves for all three materials. The integral of $I_p(T)$ over time gives the reproducible value of $P$($T$). The polarization ($P$) with $T$ for opposite $E$ are depicted in Figs. \ref{pyro}(a), \ref{pyro}(d), and \ref{pyro}(g) for EBCO, DBCO, and SBCO, respectively. The reversal of $P(T)$ due to a change in sign of $E$ ($\pm$ 4 kV/cm) signifies the ferroelectric behavior of the compounds. Importantly, the values of $P$ in the current investigation are $\sim$ 1410, $\sim$ 992, and $\sim$ 980 $\mu$C/m$^2$ for EBCO, DBCO, and SBCO, respectively for $E$ = 4 kV/cm. The $P$-values of EBCO and DBCO are $\sim$ 10 and $\sim$ 245 times higher than the reported values of $P$ for Er$_2$BaNiO$_5$ and Dy$_2$BaNiO$_5$, \cite{basu1,basu2} whereas the $P$-value of Sm$_2$BaCuO$_5$ is quite close the the reported $P$-value of Sm$_2$BaNiO$_5$.\cite{indra} 

\begin{figure}[t]
\centering
\includegraphics[width = \columnwidth]{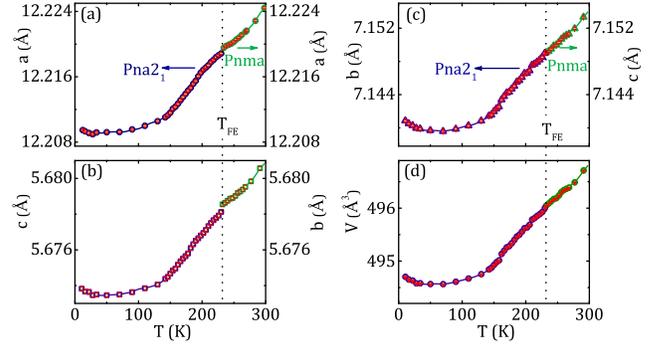}
\caption {The $T$ variations of (a) $a$, (b) $b$, (c) $c$, and (d) unit cell volume (V) for DBCO. The vertical broken line shows the $T_{FE}$.} 
\label{latt}
\end{figure}

In order to investigate the possible ME coupling, the $I_p(T)$ curves are recorded in different conditions of magnetic FC effects after cooling from $T > T_{N1}$ or $T_{N2}$, separately. The magnetic FC effects are applied with three different conditions: 1) ZFC, 2) FC protocols as described in the experimental section, and 3) cooling in a non-zero $H$ and measurements of $I_p(T)$ in the warming mode in zero magnetic field. In the above three cases the magnetic field cooling temperature and electric field poling temperature are considered same. Intriguingly, a definite signature of the peak in the $I_p(T)$ curve, pointing additional change in $P$ ($\Delta P$), is always noted for case 2) i.e. in the typical magnetic FC protocol only. In other two cases the peak in the $I_p(T)$ curves is always absent for the three samples. In addition, during record of $I_p(T)$ curve with FC protocol as described in case 2), the cooling from $T > T_{N1}$ as well as $T_{N2}$ provides the same result for EBCO and DBCO. This confirms that the R ordering at $T_{N2}$ is the key for the occurrence of magnetic FC driven $\Delta P$, whereas Cu ordering at $T_{N1}$ does not influence on the occurrence of ferroelectricity. The detailed results of the $I_p(T)$ curves for different magnetic field cooling processes are described in Fig. S2 of the Supplemental Material for DBCO. 

\begin{figure*}[t]
\centering
\includegraphics[width = 14cm]{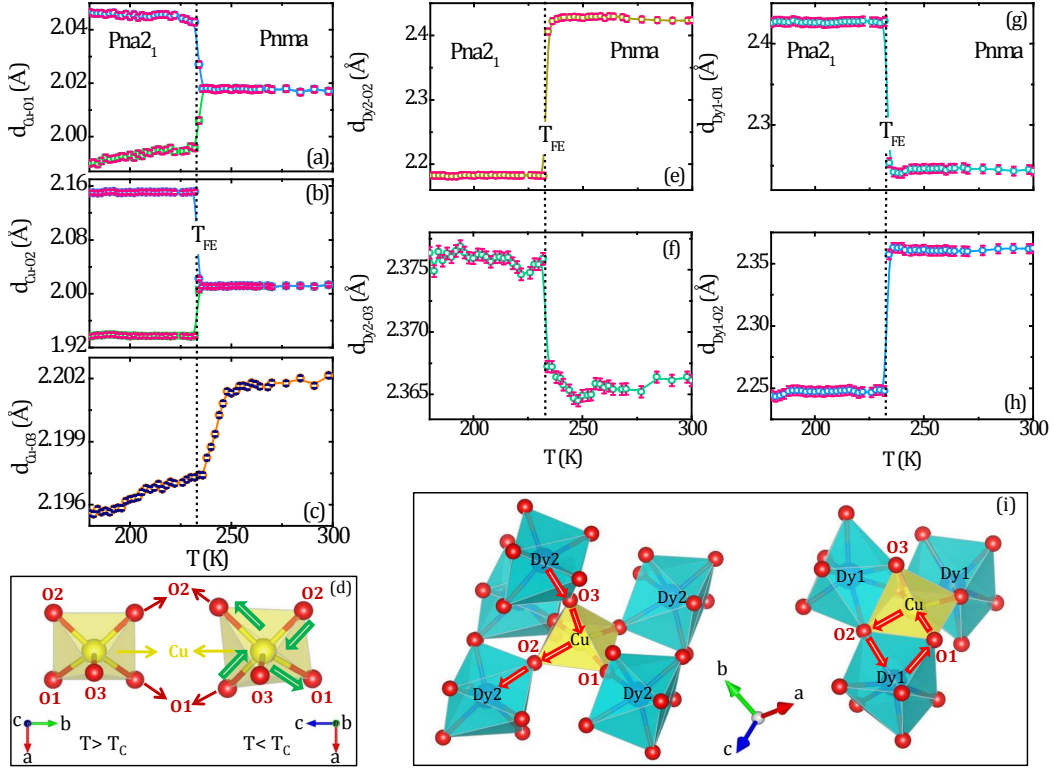}
\caption {Variations of (a) Cu-O1 ($d_{Cu-O1}$), (b) Cu-O2 ($d_{Cu-O2}$), (c) Cu-O3 ($d_{Cu-O3}$), (e) Dy2-O2 ($d_{Dy2-O2}$), (f) Dy2-O3 ($d_{Dy2-O3}$), (g) Dy1-O1 ($d_{Dy1-O1}$), and (h) Dy1-O2 ($d_{Dy1-O2}$) bond lengths with $T$ for DBCO. (d) Distortion of tetragonal $CuO_5$ pyramid below and above $T_{FE}$. (i) Distortions of the connecting DyO$_7$ and CuO$_5$ polyhedra below $T_{FE}$ involving Dy1 (right panel) and Dy2 (left panel).} 
\label{bond}
\end{figure*}

The results on the occurrence of $\Delta P$ due to magnetic FC effect are depicted in Figs. \ref{pyro}(b) and \ref{pyro}(c) for EBCO, Figs. \ref{pyro}(e) and \ref{pyro}(f) for DBCO, and Figs. \ref{pyro}(h) and \ref{pyro}(i) for SBCO. In Fig. \ref{pyro}(b) the $I_p(T)$ curves are depicted for different cooling fields ($H_{cool}$). The peaks in the $I_p(T)$ curves are observed below $T_{N2}$ of EBCO and are strongly influenced by $H_{cool}$. The peak observed in the $I_p(T)$ curves shifts with $H_{cool}$ and the magnitude is strongly dependent on $H_{cool}$. The $\Delta P(T)$ curves, as obtained from the time integration of $I_p(T)$ curves for $E$ = $\pm$ 4 kV/cm and $H_{cool}$ = 50 kOe, are depicted in the inset of Fig. \ref{pyro}(b). The reversal of $\Delta P$ due to opposite poling field ensures the ferroelectric nature of the magnetic FC driven occurrence of $\Delta P$. The $\Delta P$ at different $H_{cool}$ are depicted in Fig. \ref{pyro}(c), where inset of the figure shows the plot of $\Delta P/P$ with $H_{cool}$, where $\Delta P$ and $P$ at 2 K are the change in polarization driven by the FC effect and the value of the electric polarization recorded in zero magnetic field, as depicted in Figs. \ref{pyro}(a),  \ref{pyro}(d), and  \ref{pyro}(g) for EBCO, DBCO, and SBCO, respectively. The plot of $\Delta P/P$ with $H_{cool}$ at 2 K exhibits similar $H$ dependence of MD for ECBO. As observed in the MD with $H$ plot, the $\Delta P/P$ plot shows a sharp rise above 7.5 kOe and is followed by a decrease displaying a 'dip', above which it increases with increasing $H_{cool}$. The results of $I_p(T)$ and $\Delta P(T)$ curves for DBCO are depicted in Figs. \ref{pyro}(e) and \ref{pyro}(f), respectively. The peaks observed in the $I_p(T)$ curve are found just below $T_{N2}$. Analogous to the results of EBCO, peak in the $I_p(T)$ curves shifts and value of $\Delta P$ changes significantly depending on $H_{cool}$. As depicted in the inset of Fig. \ref{pyro}(f), the $\Delta P/P$ with $H_{cool}$ at 2 K exhibits similar behavior of the MD vs $H$ plot for DBCO. A sharp increase in $\Delta P/P$ is observed close to $\sim$ 20 kOe, analogous to that observed magnetic field driven transition at 20 kOe in the magnetization curve at 2 K. The $\Delta P/P$ shows a decrease with further increase in $H_{cool}$. The results of SBCO are different from the results of EBCO and DBCO. A sharp peak in the $I_p(T)$ curve is observed at $T_{N1}$, which is missing for EBCO and DBCO. In addition, another peak in the $I_p(T)$ curve is observed, which is opposite in direction below $T_{N2}$. Both the peaks in the $I_p$ curve remain at the same position for all the values of $H_{cool}$. The plot of $\Delta P/P$ with $H_{cool}$ shows nearly linear dependence at 2 K, as also observed in the magnetization curve as well as MD vs $H$ plot in the high field region for SBCO.  


To find out the origin of ferroelectricity at much higher temperatures than the magnetic ordering temperatures, the structural properties are investigated by the synchrotron diffraction studies over a temperature range of 10-300 K for DBCO, as a representative of isostructural R$_2$BaCuO$_5$ series. Example of a diffraction pattern together with the refinement with a $Pnma$ space group at 300 K is shown in Fig. \ref{xrd}(a) with the lattice constants, $a$ = 12.2243(9), $b$ = 5.6808(4), and $c$ = 7.1539(4) \AA. The reliability parameters of the refinement are $R_w$(\%) $\sim$ 2.30, $R_{exp}$(\%) $\sim$ 1.92, and $\sigma$ $\sim$ 0.023 at 300 K. Figure \ref{xrd}(b) shows the diffraction patterns of DBCO above $T_{FE}$ at 300 K and below $T_{FE}$ at 200 K together in a same plot. The inset of Fig. \ref{xrd}(b) highlights the change of intensity pattern of the (421) diffraction peak associated with a shift of the peak position. The changes of the (421) diffraction peak position are depicted in Fig. \ref{xrd}(c) from 240 K to 226 K with the small temperature intervals. Here, the (421) peak at different temperatures is vertically shifted for the clarification of the small changes of the peak position. Temperature variation of the integrated intensity of the (421) diffraction peak is depicted in Fig. \ref{xrd}(d), which displays an evident signature near $T_{FE}$. This signature around $T_{FE}$ in the integrated intensity plot is correlated with the change in scattering cross section and may point to a possible structural transition. The change in the integrated intensity is similar to that observed for the reported ferroelectric materials, where ferroelectricity appeared due to structural transition.\cite{indra,indra1,indra2,dey} With further decreasing temperature a sharp fall is observed around $T_{N1}$, as depicted in Fig.  \ref{xrd}(d), which indicates a strong signature of the magnetoelastic coupling. The possible occurrence of the magnetoelastic coupling at $T_{N2}$ is beyond the scope of our synchrotron data, which is recorded up to 10 K.

The diffraction patterns are refined with the high-temperature $Pnma$ space group in the entire recorded temperature range. We note that the refinement is not satisfactory below $T_{FE}$ using $Pnma$ space group, as depicted in Fig. \ref{xrd}(e). Thus a structural transformation from $Pnma$ to a polar structure is proposed for justifying the occurrence of a polar order. We use AMPLIMODE\cite{ampli} and ISODISTORT\cite{iso} softwares to find out possible noncentrosymmetric space groups having a polar structure below $T_{FE}$. Among all possible noncentrosymmetric structures the $Pna2_1$ space group has the highest symmetry, which is also a polar structure. The best fit is realized for the $Pna2_1$ space group with the lattice constants, $a$ = 12.2192(2), $b$ = 7.1492(2), and $c$ = 5.6783(6) \AA. Example of a satisfactory fit with the $Pna2_1$ space group at 200 K is shown in Fig. \ref{xrd}(f). Insets of the Figs. \ref{xrd}(e) and \ref{xrd}(f) clearly demonstrate the refinements in a small 2$\theta$ range and authenticate the better fit of the diffraction pattern using $Pna2_1$ than the $Pnma$ space group with the small reliability parameters, $R_w$(\%) $\sim$ 2.15, $R_{exp}$(\%) $\sim$ 1.91, and $\sigma$ $\sim$ 0.028. The bars below the diffraction patterns represent the diffraction peak positions and the difference plots are shown at the bottom for all the refinements. The difference plots shown at the bottom confirm a single phase without any trace amount of impurity.

Thermal variations of the lattice parameters, $a$, $b$, and $c$, as obtained from the refinements, are depicted in Figs. \ref{latt}(a), \ref{latt}(b), and \ref{latt}(c), respectively. The $T_{FE}$ is shown by the vertical broken line in the figures.  As depicted in Figs. \ref{latt}(b), and \ref{latt}(c), the $b$ and $c$ axes in $Pnma$ become $c$ and $b$ axes in $Pna2_1$ structure. An anomaly around $T_{FE}$ is observed in $a$($T$) and $b$($T$), which is not so apparent in $c$($T$). As depicted in Fig. \ref{latt}(d), any significant change in the unit cell volume is absent around $T_{FE}$. The results are analogous to that observed in few members of the orthorhombic RCrO$_3$ series, which exhibited the FE order involving structural transition.\cite{indra1,ghosh} The results further indicate that the ferroelectricity in DBCO is correlated to this structural transition from the centrosymmetric $Pnma$ to a polar $Pna2_1$ structure. We note that the structural transition involves with the considerable deformations of CuO$_5$ and DyO$_7$ polyhedra. In order to probe these deformations microscopically, the bond lengths and bond angles between different atoms are investigated further.


In the tetragonal CuO$_5$ pyramids one oxygen (O3) occupies the apex position and out of four basal oxygens two are defined as O1 and the rest two as O2 in both the $Pnma$ and $Pna2_1$ space groups. The positions of oxygen are shown in the Fig. \ref{bond}(d). The thermal variations of two Cu-O1 ($d_{Cu-O1}$) and two Cu-O2 ($d_{Cu-O2}$) bond lengths are depicted in Figs. \ref{bond}(a) and \ref{bond}(b), respectively. The step-like changes are observed in $d_{Cu-O1}$ and $d_{Cu-O2}$ around $T_{FE}$. The step-like increase and  decrease of $\sim$ 1.2 \% and 1.09 \% are noted for $d_{Cu-O1}(T)$, whereas the step-like more stronger increase and decrease of $\sim$ 7.02 \% and 3.68 \% are noted for $d_{Cu-O2}(T)$. The decrease in $d_{Cu-O3}(T)$ is found to be small as $\sim$ 0.18 \%, as depicted in Fig. \ref{bond}(c). These results indicate that distortion of tetragonal CuO$_5$ pyramid is significant in the basal plane around $T_{FE}$. The distortions of CuO$_5$ polyhedra for structural change from $Pnma$ to $Pna2_1$ space group are depicted in the right panel of Fig. \ref{bond}(d) by the arrows. 

\begin{figure}[t]
\centering
\includegraphics[width = \columnwidth]{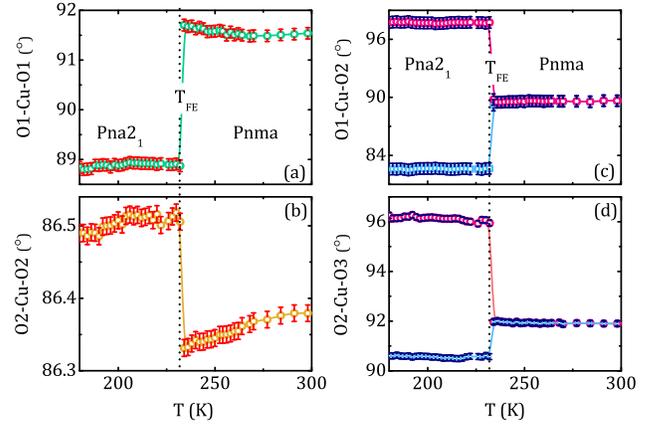}
\caption{Variations of (a) O1-Cu-O1, (b) O2-Cu-O2, (c) O1-Cu-O2, and (d) O2-Cu-O3 bond angles with $T$ for DBCO.} 
\label{ang}
\end{figure}

As observed in Fig. \ref{str}, the CuO$_5$ polyhedra are connected to each other via DyO$_7$ polyhedra, so the changes in CuO$_5$ polyhedra must influence the neighboring DyO$_7$ polyhedra. Two nonequivalent sites of Dy are defined as Dy1 and Dy2 in this article. The Dy1O$_7$ and Dy2O$_7$ prisms are joined by a common face and edge into a Dy1Dy2O$_{11}$ unit. These units linked by the common edges and faces with CuO$_5$ polyhedra form a three-dimensional network. The linkage of CuO$_5$ polyhedra with Dy1O$_7$ and Dy2O$_7$ are shown separately in the right and left panels, respectively, of Fig. \ref{bond}(i). A tetragonal CuO$_5$ pyramid is connected with five Dy2O$_7$ polyhedra through five oxygen at the edges. The thermal variations of Dy2-O2 ($d_{Dy2-O2}$) and Dy2-O3 ($d_{Dy2-O3}$) bond lengths are depicted in Figs. \ref{bond}(e) and \ref{bond}(f), respectively. In contrast to the decrease in $d_{Cu-O3}(T)$ [Fig. \ref{bond}(c)], an increase in $d_{Dy2-O3}(T)$ is observed around $T_{FE}$. Similarly, the observed decrease in $d_{Dy2-O2}(T)$ is also correlated with the increase of connected $d_{Cu-O2}(T)$ [Fig. \ref{bond}(b)]. On the other hand, the three Dy1O$_7$ polyhedra are connected by the faces with tetragonal CuO$_5$ pyramid. The thermal variations of Dy1-O1 ($d_{Dy1-O1}$) and Dy1-O2 ($d_{Dy1-O2}$) bond lengths, depicted in Figs. \ref{bond}(g) and \ref{bond}(h), also show the similar contrasting nature with the variation of connecting $d_{Cu-O1}(T)$ and $d_{Cu-O2}(T)$ around $T_{FE}$. The changes in aforesaid bond lengths are schematically represented in Fig. \ref{bond}(i) by the arrows. These results clearly imply that the deformations of both Dy1O$_7$ and Dy2O$_7$ polyhedra are strongly influenced by the distortion of CuO$_5$ polyhedra around FE order in DBCO. The results clearly infer that the emergence of electric polarization involves large distortions of the CuO$_5$ and  DyO$_7$ polyhedra. 


The thermal variations of bond angles between Cu and three O atoms are depicted in Figs. \ref{ang}(a-d). As depicted in Figs. \ref{ang}(a) and \ref{ang}(b), a sharp decrease of $\sim$ 3.03 \% for O1-Cu-O1 bond angle and the small increase of $\sim$ 0.19 \% for O2-Cu-O2 bond angle are observed around $T_{FE}$. The decrease and increase are observed in O1-Cu-O2 and O2-Cu-O3 bond angles, respectively, around $T_{FE}$ in DBCO. The maximum values of changes in O1-Cu-O2 and O2-Cu-O3 below $T_{FE}$ are remarkable as $\sim$ 9\% (increase), $\sim$ 8 \% (decrease), and  $\sim$ 4.2\% (increase), $\sim$ 1.5 \% (decrease), respectively. Our micro-structural studies indicate that the distortions of CuO$_5$ polyhedra involving the structural changes lead to a key role for the ferroelectric order in DBCO, which is analogous to that observed for RCrO$_3$ (R = Sm and Ho).\cite{indra1,ghosh} 

The structural studies propose that the ferroelctricity involves the structural transition at high temperature, which is much above the magnetic order. The results are rather analogous to that observed in few member of spin-chain compounds R$_2$BaNiO$_5$, such as, Er$_2$BaNiO$_5$\cite{basu1} and Sm$_2$BaNiO$_5$.\cite{indra} For Er$_2$BaNiO$_5$ a short range order driven ferroelectricity was proposed, whereas a structural transition to a non-centrosymmetric polar structure has been proposed for the occurrence of ferroelectricity in case of Sm$_2$BaNiO$_5$. Similar results with a much higher $T_{FE}$ than the magnetic order was observed for RCrO$_4$ series (Sm, Gd, Ho), which was proposed to be correlated with the structural transition.\cite{indra1} Furthermore, the structural distortion has been suggested for tuning $P$ value in several occasions of films\cite{shim_film,dubo_film,liu_film,dau_film,kim_film,lee_film} as well as polycrystalline compounds.\cite{indra1,zhao_str,shim_str,dey_str} The influence of structural distortion for the large polarization in bulk perovskite systems was also addressed from the first principles calculations.\cite{zhao_str} The structural distortion driven increase of $P$ has been proposed for ABi$_2$Ta$_2$O$_9$ series,\cite{shim_str} Ca$_{1−x}$La$_x$BaCo$_4$O$_7$ ($x \leq$ 0.05),\cite{dey_str} and RCrO$_4$ series.\cite{indra1} Herein, we note the $P$ value is systematic with the ionic radius of R$^{3+}$ in R$_2$BaCuO$_5$, where low ionic radius involves a high value of $P$ for EBCO. With the increase in ionic radius, the $P$ value decreases drastically for DBCO and it continues to decrease slowly for SBCO. The lower ionic radius of Er$^{3+}$ might be correlated to the larger structural distortion and thus provides the larger $P$ value for EBCO. In order to establish it, further investigations are suggested on the structural properties of other members of R$_2$BaCuO$_5$ series below $T_{FE}$.    

The magnetic FC effect driven occurrence of $\Delta P$ is another interesting results in the current investigation. In fact, the $\Delta P$ reverses its sign due to change in sign of poling field and proposes the ferroelectric nature of the appearance of $\Delta P$. Here, we emphasize on the fact that additional increase of polarization occurs only in case of magnetic FC effect, which is absent for the ZFC effect. The contrast results are observed for SBCO compared to the results for the rest two members, where the $\Delta P$ occurs below the R$^{3+}$ ordering for EBCO and DBCO and below the Cu$^{2+}$ ordering temperature for SBCO. The results may be correlated to the important observations, where the moments of Sm$^{3+}$ and Cu$^{2+}$ ions are comparable for SBCO and considerably larger moments of Er$^{3+}$ and Dy$^{3+}$ ions are noticed than the Cu$^{2+}$ moment for EBCO and DBCO. The 3$d$-4$f$ hybridization holds the key for the contrast results. After the magnetic FC process, the possible modification of the magnetic structure, as determined by the neutron diffraction studies, can elucidate on the occurrence of $\Delta P$, which is ferroelectric in nature for R$_2$BaCuO$_5$. 

In conclusion, the studies on R$_2$BaCuO$_5$ (R = Er, Dy, Sm) series reveal two important observations. 1) The occurrence of ferroelectric order associated with the higher values of electric polarization close to room temperature for EBCO and DBCO. Although the ferroelectric order emerges for SBCO at slightly lower temperature than the $T_{FE}$ of EBCO and DBCO, the $T_{FE}$ in all the members is much higher than the corresponding magnetic ordering temperatures. A structural transition to the polar structure with a $Pna2_1$ space group correlates the ferroelectric order. 2) The unusual magnetoelectric consequences are observed for all three members of R$_2$BaCuO$_5$. Intriguingly, an additional increase of polarization, which is ferroelectric in nature, appears only in case of magnetic field-cooled effect. The results propose that the R$_2$BaCuO$_5$ series is a new family of multiferroics.  

\vspace{0.2in}
\noindent
{\bf Acknowledgment}\\
S.G. acknowledges SERB, India project (Project No. SB/S2/CMP-029/2014) for the financial support. S.G. also acknowledges DST, India for the financial support to perform experiment at PETRA III, DESY, Germany for synchrotron diffraction studies (Proposal No. I-20150193).

\vspace{0.5in}

\textbf{Supplemental Material- High-temperature ferroelectric order and magnetic field-cooled effect driven magnetoelectric coupling in R$_2$BaCuO$_5$ (R= Er, Dy, Sm)}


\begin{figure}[h!]
\centering
\includegraphics[width = 0.6\columnwidth]{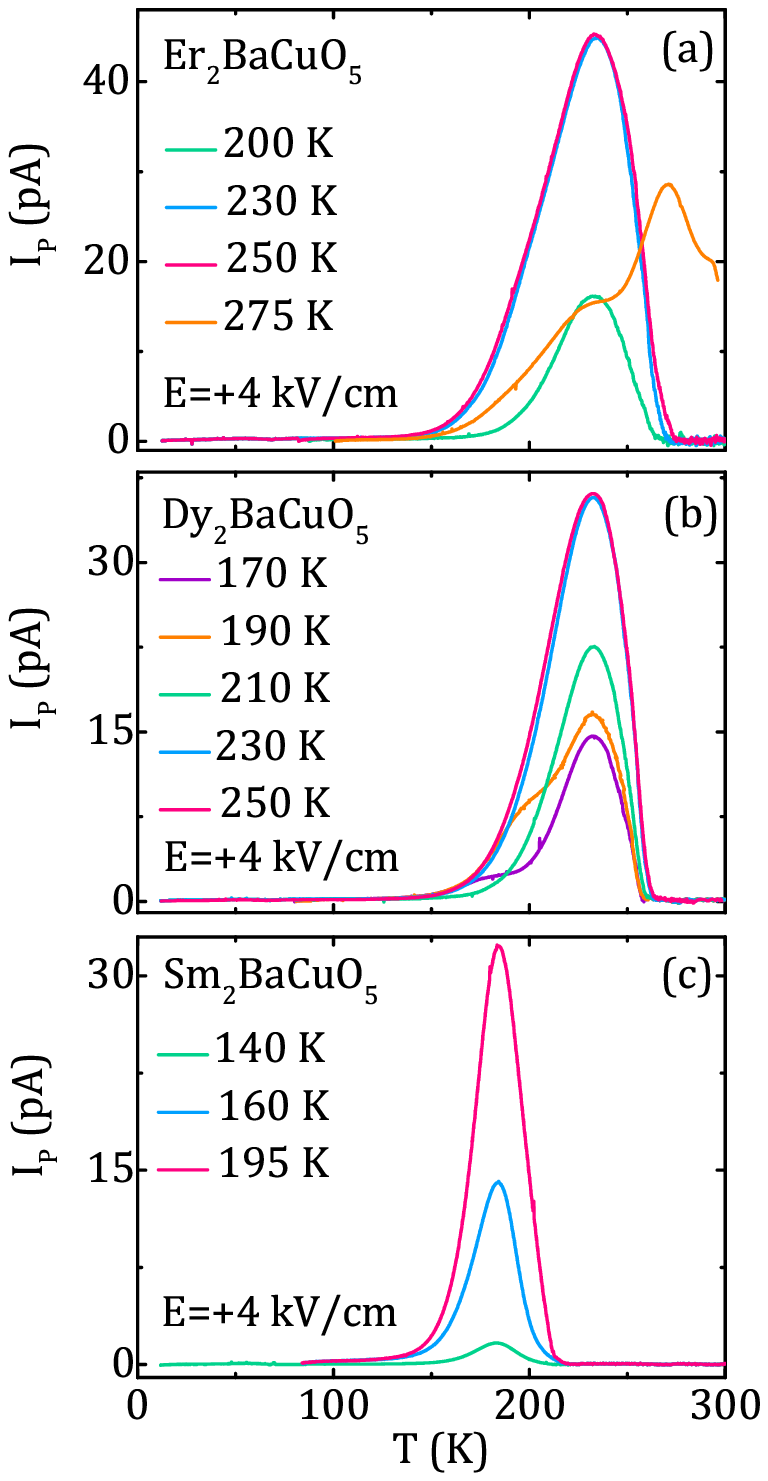}
\captionsetup{labelformat=empty}
\caption*{FIG. S1: Temperature variations of pyroelectric current ($I_P$) for different poling temperatures at poling field ($E$) of 4 kV/cm for (a) EBCO, (b) DBCO, and (c) SBCO.} 
\label{Tpole}
\end{figure}

In order to confirm the peaks in $I_p(T)$ at high temperatures appear due to genuine occurrence of polar order for all three compounds, the pyroelectric current is recorded at different poling temperatures ($T_{pole}$) for a 4 kV/cm poling field. Here, samples are always cooled from the selected $T_{pole}$ down to 2 K and $I_P(T)$ is measured during warming mode in zero electric field. The results of $I_P(T)$ at different $T_{pole}$ are depicted in Figs. \ref{Tpole}(a), \ref{Tpole}(b), and \ref{Tpole}(c) for  EBCO, DBCO, and SBCO, respectively. Here, different values of $T_{pole}$ are selected below, above, and close to the high temperature peak of the respective sample.

 In case of EBCO, an additional peak above $\sim$ 235 K can be clearly observed at $\sim$ 270 K when poling is done at 275 K. The high temperature peak is absent when the poling is done at temperature 250 K. This confirms that the high-temperature large peak appears due to the extrinsic thermally stimulated depolarization currents (TSDC). The peak height around 235 K is nearly same for poling at 230 K and 250 K. $T_{pole}$=200 K reduces the peak height as we observe in Fig. \ref{Tpole}(a). Similarly for DBCO and SBCO an apparent peak is always observed around $\sim$ 232 K and $\sim$ 184 K for different values of $T_{pole}$, as depicted in the Fig. \ref{Tpole}(b) and \ref{Tpole}(c), respectively. As shown in Fig. \ref{Tpole}(b), the peak height around 232 K is almost same for $T_{pole}$= 230 K and 250 K, and reduces gradually for $T_{pole}$= 210 K, 190 K, and 170 K for DBCO. These results of DBCO are similar as EBCO. Fig. \ref{Tpole}(c) also shows that the peak height around 184 K also reduces gradually for $T_{pole}$= 195 K, 160 K, and 140 K for SBCO. The peak due to extrinsic thermally stimulated depolarization currents (TSDC) is not observed for DBCO and SBCO when the poling is done from much higher temperature. These results are consistent with those previously reported for CuCrO$_2$\cite{ngo} and Fe-dope SrTiO$_3$\cite{liu}. So the appearance of a peak around $\sim$ 235 K, 232 K, and 184 K for EBCO, DBCO, and SBCO, respectively, in the pyroelectric current measurement indicate the genuine occurrence of the ferroelectricity in these three compounds.\cite{chen} 

\begin{figure}[h!]
\centering
\includegraphics[width = 0.8\columnwidth]{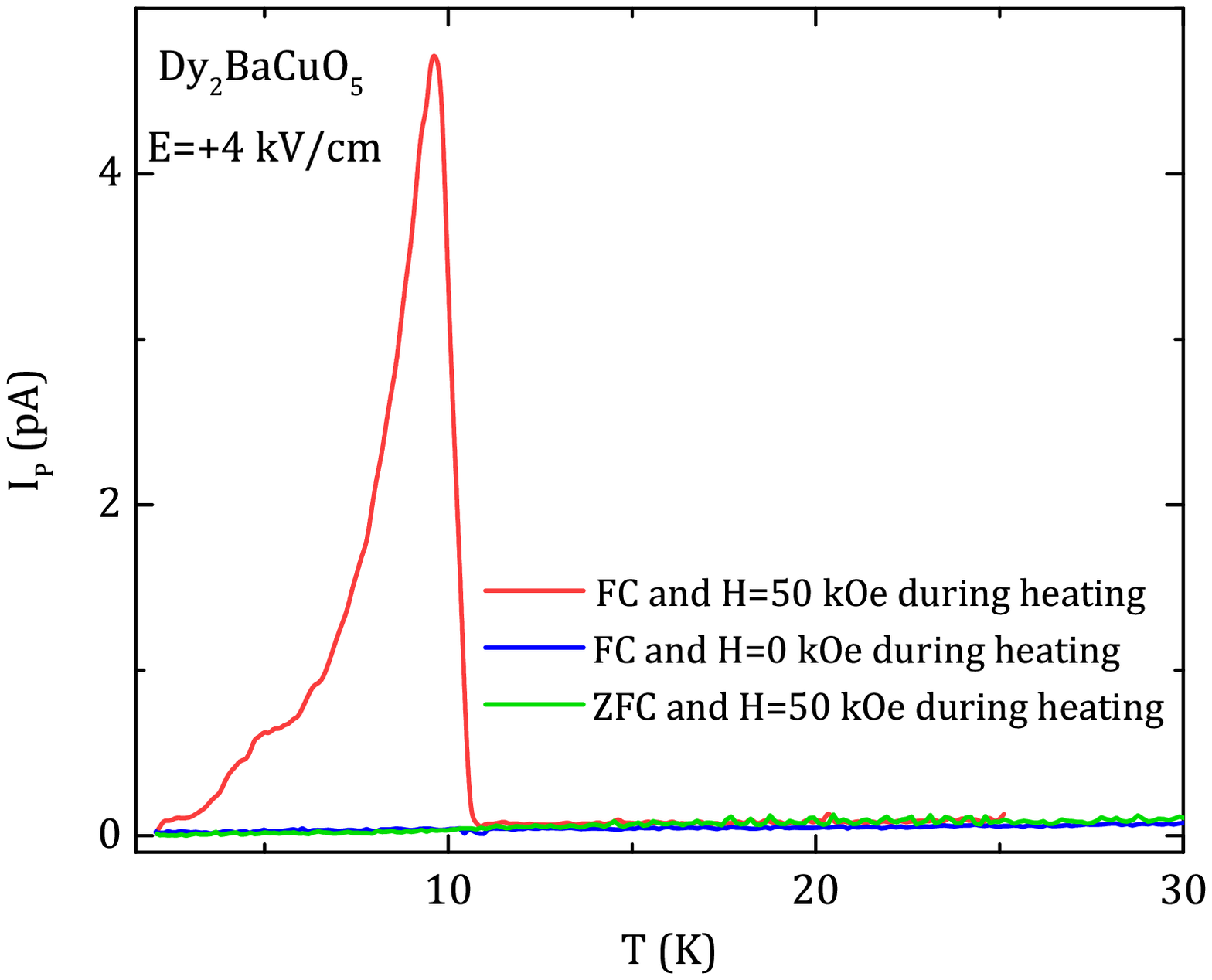}
\captionsetup{labelformat=empty}
\caption*{FIG. S2: Temperature dependence of $I_P$ curves of DBCO recorded in different conditions of magnetic FC effects.} 
\label{FC}
\end{figure}

In order to investigate the possible ME coupling, the $I_P(T)$ curves are recorded in different conditions of magnetic FC effects after cooling from temperature above $T_{N1}$ or $T_{N2}$, separately. The results of different magnetic FC effects after cooling from $T$ greater than $T_{N1}$ for DBCO are shown in the Fig. \ref{FC}. The pyroelectric current is recorded for $E$=4 kV/cm poling field. Here the sample is always cooled down to 2 K and $I_P$ is measured during warming mode in zero electric field. As depicted in Fig. \ref{FC}, the magnetic field of 50 kOe are applied with three different conditions, which are indicated by three different $I_P(T)$ curves. Green-curve shows ZFC (H=0 kOe during cooling) and measurement of $I_P(T)$ in the warming mode with H=50 kOe, blue-curve shows FC (H=50 kOe during cooling) and measurement of $I_P(T)$ in the warming mode in zero magnetic field, and red-curve shows FC (H=50 kOe during cooling) and measurement of $I_P(T)$ in the warming mode with H=50 kOe. In the above three cases the magnetic field cooling temperature and electric field poling temperature are considered same. Intriguingly, the definite signature of the peak in the $I_P(T)$ curve below $T_{N2}$ is observed for FC and H=50 kOe during heating, i.e. in the typical magnetic FC protocol only.

\end{document}